\documentclass[aps,prd,twocolumn,superscriptaddress,showpacs,amsmath,amssymb]{revtex4}

\usepackage{graphicx}
\usepackage{dcolumn}
\usepackage{bm}

\usepackage{amssymb,amsthm,amsmath}
\usepackage{graphicx}
\usepackage{dcolumn}
\usepackage{bm}
\usepackage{setspace}
\usepackage{enumerate}

\newcommand{\wzgqq}{\gamma{}(W/Z)\rightarrow{}\gamma{}q\bar{q}\ }
\newcommand{\ptg}{p_T^\gamma }

\newcommand{\dps}{{\tt PHOTON\_DIJET\_L2\_DPS}\ }

\newcommand{\eph}{\epsilon_{ph}}
\newcommand{\ejets}{\epsilon_{jet}}
\newcommand{\ee}{\epsilon}
\newcommand{\nno}{N_{OUT}}
\newcommand{\etal}{{\em et al.}}
\newcommand{\lh}{{\mathcal L}}
\newcommand{\nh}{{\mathcal N}}

\begin{document}

\title{
Search for Hadronic Decays of  W and Z Bosons in Photon Events in $p\bar p$ Collisions at $\sqrt{s}$=1.96 GeV}

\affiliation{Institute of Physics, Academia Sinica, Taipei, Taiwan 11529, Republic of China} 
\affiliation{Argonne National Laboratory, Argonne, Illinois 60439} 
\affiliation{Institut de Fisica d'Altes Energies, Universitat Autonoma de Barcelona, E-08193, Bellaterra (Barcelona), Spain} 
\affiliation{Baylor University, Waco, Texas  76798} 
\affiliation{Istituto Nazionale di Fisica Nucleare, University of Bologna, I-40127 Bologna, Italy} 
\affiliation{Brandeis University, Waltham, Massachusetts 02254} 
\affiliation{University of California, Davis, Davis, California  95616} 
\affiliation{University of California, Los Angeles, Los Angeles, California  90024} 
\affiliation{University of California, San Diego, La Jolla, California  92093} 
\affiliation{University of California, Santa Barbara, Santa Barbara, California 93106} 
\affiliation{Instituto de Fisica de Cantabria, CSIC-University of Cantabria, 39005 Santander, Spain} 
\affiliation{Carnegie Mellon University, Pittsburgh, PA  15213} 
\affiliation{Enrico Fermi Institute, University of Chicago, Chicago, Illinois 60637} 
\affiliation{Comenius University, 842 48 Bratislava, Slovakia; Institute of Experimental Physics, 040 01 Kosice, Slovakia} 
\affiliation{Joint Institute for Nuclear Research, RU-141980 Dubna, Russia} 
\affiliation{Duke University, Durham, North Carolina  27708} 
\affiliation{Fermi National Accelerator Laboratory, Batavia, Illinois 60510} 
\affiliation{University of Florida, Gainesville, Florida  32611} 
\affiliation{Laboratori Nazionali di Frascati, Istituto Nazionale di Fisica Nucleare, I-00044 Frascati, Italy} 
\affiliation{University of Geneva, CH-1211 Geneva 4, Switzerland} 
\affiliation{Glasgow University, Glasgow G12 8QQ, United Kingdom} 
\affiliation{Harvard University, Cambridge, Massachusetts 02138} 
\affiliation{Division of High Energy Physics, Department of Physics, University of Helsinki and Helsinki Institute of Physics, FIN-00014, Helsinki, Finland} 
\affiliation{University of Illinois, Urbana, Illinois 61801} 
\affiliation{The Johns Hopkins University, Baltimore, Maryland 21218} 
\affiliation{Institut f\"{u}r Experimentelle Kernphysik, Universit\"{a}t Karlsruhe, 76128 Karlsruhe, Germany} 
\affiliation{Center for High Energy Physics: Kyungpook National University, Daegu 702-701, Korea; Seoul National University, Seoul 151-742, Korea; Sungkyunkwan University, Suwon 440-746, Korea; Korea Institute of Science and Technology Information, Daejeon, 305-806, Korea; Chonnam National University, Gwangju, 500-757, Korea} 
\affiliation{Ernest Orlando Lawrence Berkeley National Laboratory, Berkeley, California 94720} 
\affiliation{University of Liverpool, Liverpool L69 7ZE, United Kingdom} 
\affiliation{University College London, London WC1E 6BT, United Kingdom} 
\affiliation{Centro de Investigaciones Energeticas Medioambientales y Tecnologicas, E-28040 Madrid, Spain} 
\affiliation{Massachusetts Institute of Technology, Cambridge, Massachusetts  02139} 
\affiliation{Institute of Particle Physics: McGill University, Montr\'{e}al, Canada H3A~2T8; and University of Toronto, Toronto, Canada M5S~1A7} 
\affiliation{University of Michigan, Ann Arbor, Michigan 48109} 
\affiliation{Michigan State University, East Lansing, Michigan  48824} 
\affiliation{University of New Mexico, Albuquerque, New Mexico 87131} 
\affiliation{Northwestern University, Evanston, Illinois  60208} 
\affiliation{The Ohio State University, Columbus, Ohio  43210} 
\affiliation{Okayama University, Okayama 700-8530, Japan} 
\affiliation{Osaka City University, Osaka 588, Japan} 
\affiliation{University of Oxford, Oxford OX1 3RH, United Kingdom} 
\affiliation{University of Padova, Istituto Nazionale di Fisica Nucleare, Sezione di Padova-Trento, I-35131 Padova, Italy} 
\affiliation{LPNHE, Universite Pierre et Marie Curie/IN2P3-CNRS, UMR7585, Paris, F-75252 France} 
\affiliation{University of Pennsylvania, Philadelphia, Pennsylvania 19104} 
\affiliation{Istituto Nazionale di Fisica Nucleare Pisa, Universities of Pisa, Siena and Scuola Normale Superiore, I-56127 Pisa, Italy} 
\affiliation{University of Pittsburgh, Pittsburgh, Pennsylvania 15260} 
\affiliation{Purdue University, West Lafayette, Indiana 47907} 
\affiliation{University of Rochester, Rochester, New York 14627} 
\affiliation{The Rockefeller University, New York, New York 10021} 
\affiliation{Istituto Nazionale di Fisica Nucleare, Sezione di Roma 1, University of Rome ``La Sapienza," I-00185 Roma, Italy} 
\affiliation{Rutgers University, Piscataway, New Jersey 08855} 
\affiliation{Texas A\&M University, College Station, Texas 77843} 
\affiliation{Istituto Nazionale di Fisica Nucleare, University of Trieste/\ Udine, Italy} 
\affiliation{University of Tsukuba, Tsukuba, Ibaraki 305, Japan} 
\affiliation{Tufts University, Medford, Massachusetts 02155} 
\affiliation{Waseda University, Tokyo 169, Japan} 
\affiliation{Wayne State University, Detroit, Michigan  48201} 
\affiliation{University of Wisconsin, Madison, Wisconsin 53706} 
\affiliation{Yale University, New Haven, Connecticut 06520} 
\author{T.~Aaltonen}
\affiliation{Division of High Energy Physics, Department of Physics, University of Helsinki and Helsinki Institute of Physics, FIN-00014, Helsinki, Finland}
\author{J.~Adelman}
\affiliation{Enrico Fermi Institute, University of Chicago, Chicago, Illinois 60637}
\author{T.~Akimoto}
\affiliation{University of Tsukuba, Tsukuba, Ibaraki 305, Japan}
\author{M.G.~Albrow}
\affiliation{Fermi National Accelerator Laboratory, Batavia, Illinois 60510}
\author{B.~\'{A}lvarez~Gonz\'{a}lez}
\affiliation{Instituto de Fisica de Cantabria, CSIC-University of Cantabria, 39005 Santander, Spain}
\author{S.~Amerio}
\affiliation{University of Padova, Istituto Nazionale di Fisica Nucleare, Sezione di Padova-Trento, I-35131 Padova, Italy}
\author{D.~Amidei}
\affiliation{University of Michigan, Ann Arbor, Michigan 48109}
\author{A.~Anastassov}
\affiliation{Rutgers University, Piscataway, New Jersey 08855}
\author{A.~Annovi}
\affiliation{Laboratori Nazionali di Frascati, Istituto Nazionale di Fisica Nucleare, I-00044 Frascati, Italy}
\author{J.~Antos}
\affiliation{Comenius University, 842 48 Bratislava, Slovakia; Institute of Experimental Physics, 040 01 Kosice, Slovakia}
% ...
\collaboration{CDF Collaboration\footnote{With visitors from $^a$University of Athens, 15784 Athens, Greece, 
$^b$Chinese Academy of Sciences, Beijing 100864, China, 
$^c$University of Bristol, Bristol BS8 1TL, United Kingdom, 
$^d$University Libre de Bruxelles, B-1050 Brussels, Belgium, 
$^e$University of California Irvine, Irvine, CA  92697, 
$^f$University of California Santa Cruz, Santa Cruz, CA  95064, 
$^g$Cornell University, Ithaca, NY  14853, 
$^h$University of Cyprus, Nicosia CY-1678, Cyprus, 
$^i$University College Dublin, Dublin 4, Ireland, 
$^j$University of Edinburgh, Edinburgh EH9 3JZ, United Kingdom, 
$^k$University of Heidelberg, D-69120 Heidelberg, Germany, 
$^l$Universidad Iberoamericana, Mexico D.F., Mexico, 
$^m$University of Manchester, Manchester M13 9PL, England, 
$^n$Nagasaki Institute of Applied Science, Nagasaki, Japan, 
$^o$University de Oviedo, E-33007 Oviedo, Spain, 
$^p$Queen Mary, University of London, London, E1 4NS, England, 
$^q$Texas Tech University, Lubbock, TX  79409, 
$^r$IFIC(CSIC-Universitat de Valencia), 46071 Valencia, Spain, 
}}
\noaffiliation

\date{\today}

\begin{abstract}
We report on a search for the process $p\bar{p}\rightarrow \gamma+W/Z$ with
$W/Z\rightarrow q\bar{q}$ in events containing two jets and a photon at the
center-of-mass energy $\sqrt{s}=1.96$ TeV, using 184 pb$^{-1}$ of data
collected by the CDF II detector. A neural network event selection has been
developed to optimize the rejection of the large QCD production background; it
is shown that this method gives a significant improvement in both
signal-to-noise ratio and signal sensitivity, as compared with an event
selection based on conventional cuts.  An upper limit is presented for the
$\gamma+W/Z$ production cross section with the $W$ and $Z$ decaying
hadronically.
\end{abstract}

\maketitle

\section{INTRODUCTION}

The identification of gauge boson hadronic decays is extremely challenging at
hadron colliders, since a small two-jet resonance needs to be extracted from a
huge QCD multi-jet background.  At the Tevatron only the favorable
circumstance of $W$'s generated in top quark decays has allowed for a
successful identification of the $W$ hadronic resonance ~\cite{top-W-had}.
Nevertheless, the ability to extract hadronic resonances submerged in a large
QCD background is of paramount importance in the search of new particles with
dominantly hadronic decays.  The most important example is the Higgs boson for
which no direct evidence has yet been observed.

At the Tevatron, one of the most promising signatures for the Higgs
observation is the associated production with a $W(Z)$, where the
Higgs decays into two jets ~\cite{Higgs-Report}. However, at the
center of mass energy $\sqrt{s}=1.96$ TeV, the standard model (SM) Higgs boson cross
section is much smaller than that for the  non-resonant
$W+jj$ production, and thus, sophisticated techniques are needed
to suppress the QCD background while maintaining a high signal
detection efficiency.


In this respect, identification of dijet resonances of the $W$ and $Z$
bosons provides an important test bench for developing such
techniques, due to the high statistical sample that can be collected
and the fact that their characteristics are well known.
In addition, a highly populated $W/Z$ boson dijet mass peak is an
excellent tool to constrain the jet energy scale and also to improve
the dijet mass resolution, two essential ingredients for precision
measurement of signatures with jets in the final state.

At hadron colliders, a mass peak from $W(Z)\rightarrow jj$ was
reconstructed in the inclusive dijet events by the UA2
collaboration~\cite{UA2} at $\sqrt{s}=630$ GeV. With a signal over
background ratio ($S/B$) of about 1/35, about 5000 events
were observed. At $\sqrt{s}=1.96$ TeV, the QCD dijet production cross
section increases by approximately a factor 35 for 20 GeV jets,
making the production rate too high to be handled by the data acquisition system.
However, this is not the case when the $W(Z)$ is produced in association with
another gauge boson ($\gamma$,$W$,$Z$).

Because the $\gamma+W(Z)$ cross section is one order of
magnitude higher than the heavy diboson production $WW+WZ$,
these events offer in principle the best opportunity to identify the $W(Z)\rightarrow
jj$ resonance. 

In addition, the diboson production with a photon is interesting in its own
right. In fact, the $\gamma+W(Z)$ production is directly correlated to the
non-Abelian character of the electroweak theory, and is sensitive to physics
beyond the standard model through enhancement of the trilinear $WW\gamma$
coupling and possible contributions of the $ZZ\gamma$ and $Z\gamma\gamma$
couplings forbidden in the standard model. Although such effects have
already been searched for in the leptonic channels of $W(Z)\gamma$ events
\cite{CDF-lepton}, the successful identification of such events also in the
hadronic channels could concur for an even more stringent test of the SM in
this sector.

\subsection{Analysis Overview}\label{sec-overview}

In this paper we report on a search for $W(Z)$ decaying into two jets based on
a sample of $\gamma+jj$ data collected with the CDF II detector between July
2003 and September 2004 ~\cite{Bocci-thesis}, corresponding to an integrated
luminosity of 184$\pm$7 pb$^{-1}$ \cite{Lum-error}.  In a previous study of
this signature performed by the CDF collaboration at $\sqrt{s}=1.8$ GeV and
using 90 pb$^{-1}$ of data \cite{CDF-RunIMarina}, a significance
($S/\sqrt{S+B}$) of 0.3 was achieved, with a S/B of about 1/100.  In the study
reported here, in addition to an improved online event selection, a neural
network based technique is employed to enhance the significance.

The expected shape of the $W(Z)$ mass distribution ($m_{jj}^{W/Z}$) is
derived from simulated SM signal events. The shape of the background
is determined directly from the data by fitting the observed dijet
mass distribution ($m_{jj}$) in the control region, {\em i.e.}
excluding the part of the $m_{jj}$ spectrum around the W/Z boson mass
value where the signal is expected to be visible (signal region).

Because of the steeply falling behavior of the $m_{jj}$ distribution, it is
important to have unbiased control regions both below and above the signal
region to obtain an accurate description of the background.  Extreme care is
taken in choosing the online and offline selection cuts in order  to not
deplete the control region at low values of $m_{jj}$. In fact, since such
region has the biggest weight in the fit, it ultimately determines the accuracy with
which the background estimate can be determined.  Such accuracy is
particularly crucial in cases with very low $S/B$ ratio, like the search reported in
this article.
The excess in
the signal region over the smooth background - if consistent with the
SM signal shape - can then be attributed to $W(Z)$ decaying into jets.

This paper is organized as follows. In
section~\ref{sec-SM-Prediction}, a description of the processes
involved in the $W(Z)\gamma$ production is provided as well as the SM
cross section predictions.  Detector and trigger descriptions follow in
Sections~\ref{sec-Detector} and~\ref{sec-Trigger}.  In
Sections~\ref{sec-EventSelection} and~\ref{sec-Systematics} event
selection criteria and expected event yield, along with their
systematic uncertainties, are outlined.  The neural network based
selection and its performance is described in
Section~\ref{sec-NN-base}. Sections~\ref{sec-Results}
and~\ref{sec-Limit} discuss the results followed by the conclusions.

\section{STANDARD MODEL PREDICTION FOR THE $W(Z)\gamma$ CROSS SECTION\label{sec-SM-Prediction}}

The tree-level Feynman diagrams for $W\gamma$ and $Z\gamma$ production are
shown in Fig. 1.  Figure 1(a) and Fig. 1(b) show the $t$-channel and $u$-channel
$W(Z)$ production respectively, where a photon is radiated from one of the
incoming quarks. Figure 1(c) and Fig. 1(d) show the processes where a photon is
radiated from the decay quarks of the $W(Z)$ boson. In these latter cases the
$W(Z)$ boson resonance cannot be reconstructed from the two-body mass of the
final quarks.  The final state of these processes is very similar from both
kinematic and topological standpoints to some components of the background
 in our sample (Sec.~\ref{sec-NN}). Because our analysis cuts have a
high background rejection power, 1(c) and 1(d) radiative decays contributions are
strongly suppressed in the sample selected. Finally the process involving the
three vector boson coupling $WW\gamma$ is shown in Fig. 1(e).


\begin{figure*}[!htb]
\begin{center}
\includegraphics[width=14cm]{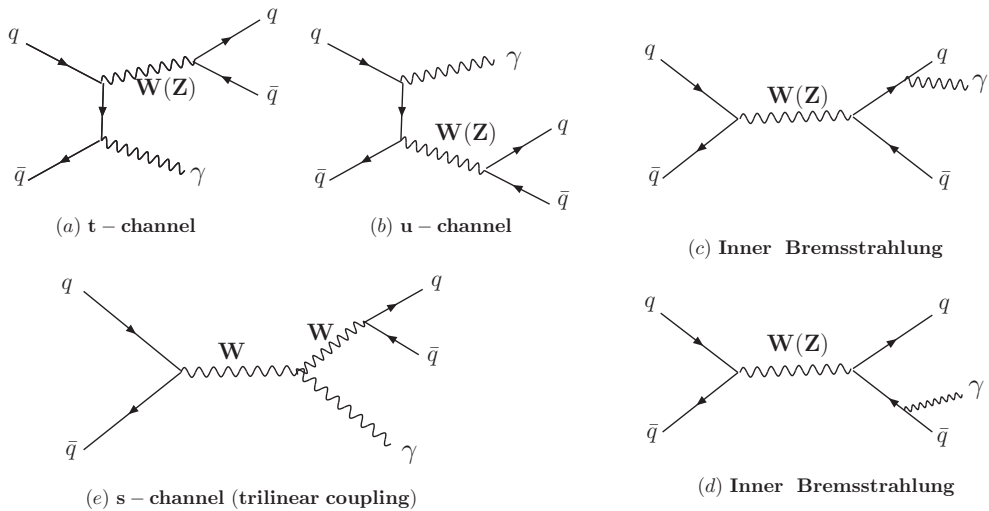}
\caption{(a,b,e) Feynman diagrams at the tree level for the process
  $q\bar{q}\rightarrow{}W(Z)\gamma\rightarrow{}q\bar{q}\gamma$. The
  s-channel for the $Z\gamma{}$ production is forbidden in the
  standard model. (c,d) Radiative W(Z) decays diagrams. A
  bremsstrahlung photon is emitted by one of the two quarks from the
  W(Z) decay.
\label{fig:wzg_graph}}
\end{center}
\end{figure*}

The $p\bar{p}\rightarrow{}W(Z)\gamma\rightarrow q \bar{q} \gamma$
predictions were determined using the {\sc pythia} \cite{Pythia} Monte
Carlo (MC) generator.  It calculates the matrix elements at leading order
(LO) and in the narrow boson-width approximation in which radiative
boson decay diagrams (Fig. 1(c) and 1(d)) are ignored. The QCD
initial/final state radiation as well as subsequent parton
fragmentation and hadronization were also provided by {\sc pythia} .
The factorization scale $Q$ was set equal to the center of mass energy
of the incoming quarks $\sqrt{\hat{s}}$. The CTEQ5L \cite{cteq5L}
parton distribution functions were used.

The {\sc pythia} calculations were compared to the predictions obtained
with {\sc madgraph} \cite{madgraph}, a tree level matrix element
calculator that, in contrast to {\sc pythia}, does not use the
narrow boson-width approximation.  The radiative contribution was
suppressed by requiring the invariant mass of the W(Z) di-quark decays to
be greater than 74(85) GeV. In addition, to avoid collinear emission
divergences the distance between the photon and the quarks in the
$\eta$-$\phi$~\cite{var-def} space was required to be greater than 0.4.  The final
state observables from the MC simulation were also compared.  The $p_T$ and
pseudorapidity distribution  of the two outgoing
partons and of the photon as well as the pair-wise separation,
defined as $\Delta R_{ij}=\sqrt{(\phi_i-\phi_j)^2+(\eta_i-\eta_j)^2}$,
$(i,j)=1,2,3,$ $i\neq j$, were in excellent agreement.  The
resulting {\sc madgraph} $W\gamma$ ($Z\gamma$) cross section is
11\% higher than the value predicted by {\sc pythia }. The
{\sc pythia} cross section prediction is scaled for $O(\alpha_S)$ QCD
contributions ($k$-factor) coming from subprocesses with either
virtual gluon loops or gluon/quark emissions in the initial state.
The magnitude of this correction, averaged over the photon spectrum in the
region $p_T^{\gamma}>10$ GeV, is 1.55 for the $W\gamma$ process
~\cite{nlo_b1} and 1.44 for the $Z\gamma$ ~\cite{nlo_b2}.  Including
this $k$-factor the SM prediction for the signal in the
kinematic region $p_T^{\gamma}>10$ GeV and $|\eta^{\gamma}|<$ 1.2 is
\begin{eqnarray*}
 \sigma_{W(Z)\gamma} &=&
 \sigma(p\bar{p}\rightarrow{}W\gamma)\times{}{\mathfrak
 B}(W\rightarrow{}q\bar{q}) \\
&+& \sigma(p\bar{p}\rightarrow{}Z\gamma)\times{}{\mathfrak
 B}(Z\rightarrow{}q\bar{q}) = 20.5 \pm 2.5 \ \ pb.
\end{eqnarray*}
The 12\% uncertainty accounts for the discrepancy between {\sc pythia}
and {\sc madgraph} cross sections (11\%), for the $k$-factor (3\%),
the factorization scale (1.5\%), and the parton distribution function
(4.8\%) uncertainties. It is interesting to notice that in contrast to
the inclusive production, where the $W$ cross section is about three
times larger than the $Z$ cross section~\cite{UA2},  for the production of the $W$
and $Z$ in association with a photon, the SM predicts similar cross
sections ($\sigma_{W\gamma}=9.9$ pb and $\sigma_{Z\gamma}=10.6$ pb).

\section{DETECTOR DESCRIPTION\label{sec-Detector}}
A detailed description of the CDF II detector can be found
elsewhere~\cite{CDF-detector}.  Here we briefly describe the aspects
of the detector relevant for this analysis. The tracking system is a
magnetic spectrometer consisting of a 90-cm long cylindrical silicon
micro-strip detector surrounded by a 3.1 m long drift chamber, both
immersed in a 1.4 T magnetic field.  The calorimeter consists of an
electromagnetic ({\sc em}) and a hadronic ({\sc had}) compartments
covering both central ($|\eta|<1.1$) and forward ($1.1<|\eta|<3.6$)
regions.  Both calorimeters are segmented into projective towers.  The
tower size in the central calorimeter is approximately
$0.11(\eta)\times 15^\circ(\phi)$, and the resolution is about
$13.5\%/\sqrt{E_T} \oplus 2\%$ for electrons (where
$E_T=E\sin{\theta}$ and $E$ is measured in GeV). Embedded in the
central calorimeter is a a multiwire proportional chamber ({\sc ces}),
located at a depth of approximately six radiation lengths where the
density of the energy deposited by an {\sc em} shower is at a maximum. Cathode
strips and anode wires, with a channel spacing between 1.5 and 2 cm,
running along the azimuthal and the beam line direction respectively
provide precise information on the electromagnetic shower centroid as
well as the shower profile in the transverse direction.  Another wire
chamber ({\sc cpr}) is located between the magnet coil and the central
calorimeter modules.  It measures the signals from early showers of
electromagnetic particles occurring in the coil.  The {\sc ces} and
{\sc cpr} systems are used to discriminate prompt photon from
multi-photon decay products of neutral mesons, $\pi^0$'s, $\eta$'s or
$K_S$'s.

The data were collected with a three level trigger system. At level 1 (L1),
a simple selection can be made based on the presence of tracks above a
fixed $p_T$ threshold, on the total energy deposited in the calorimeter, or
on single calorimeter trigger tower energies (a trigger tower consists
of two calorimeter towers adjacent  in the $z$ direction).
At level 2 (L2), custom built hardware is used to reconstruct calorimeter
energy clusters, apply isolation requirements for photons and
electrons, identify muons, and measure track displacements from the
primary vertex.
At level 3 (L3), events are fully reconstructed with the same algorithms
used in the offline analysis.
The transverse energies however are calculated using the nominal interaction
point, instead of the actual event vertex position.  
\section{TRIGGER SELECTION\label{sec-Trigger}}
In the analysis reported in this article the main source of background
is the non-resonant QCD $\gamma+jj$ production. In addition, a large
contribution from three-jet production is also expected. Both of these
background processes have rather large event rates.  As a consequence,
an elaborate triggering scheme is needed to reduce their rate to
levels that can be handled by the current data acquisition
hardware. The main challenge is to keep the photon $p_T$ threshold low
enough in order not to bias significantly the data $m_{jj}$ distribution
 below the signal region. Only with this requirement can an
accurate determination of the background shape be successfully
carried out (see Sec.~\ref{sec-overview}).  However, an inclusive
photon trigger with a low $p_T^{\gamma}$ threshold results in an
unacceptably large rate.  We designed
a trigger taking into account the above constraints. Details of the
trigger specifications are outlined in the following sections.

\subsection{Level 1 and level 2 Selection}
At level 1 events with a trigger tower with $E_T>8$ GeV and at least
89\% of its energy deposited in the {\sc em} section are selected.  At
level 2, electromagnetic clusters are reconstructed combining towers
with $E_T>7.5$ GeV adjacent to a seed tower. A seed tower must have an
$E_T>8$ GeV with 89\% of its energy deposited in the {\sc em}
calorimeter.  Only {\sc em} clusters with $E_T>12$ GeV and isolated
from other deposits of energy are selected. The isolation requirement
proceeds as follows. The sum of the transverse energies is determined
in a) 8 towers surrounding the seed tower and b) all four combinations
of ten towers in a $4\times 3$ region surrounding the
seed and one adjacent tower. The lowest of these five sums is required
to be less than 1~GeV. Such a strict isolation requirement provides
significant rejection against the high-rate neutral meson
multi-photon-decay background and against photons radiated by quarks
or gluons.  To further reduce the L2 output rate the presence of a
significant hadronic activity were added on top of the photon
requirement. The L2 hardware jet finder was exploited to identify
clusters of energetic towers where a nearest neighbor algorithm
with a seed tower threshold of 3 GeV is used.
The trigger requires the presence of at least two such L2 clusters, one of
which corresponds to the photon, with the seed in the region $|\eta|<1.78$. 
To maximize efficiency for low Et jets we apply no explicit
requirement on cluster energy.
Instead, the total transverse energy of the calorimeter trigger towers
$\Sigma E_T$, excluding the photon candidate energy, is required to
be greater than 20 GeV.  The trigger rate reduction brought about by
these extra cuts allows the photon $E_T$ threshold to be set as low as
12 GeV.
\subsection{Level 3 Selection}
At level 3, {\sc em} clusters are formed by combining towers with more
than 2 GeV of energy with their two nearest neighbors in
pseudorapidity. Only clusters with 95\% of their energy in the
electromagnetic calorimeter are selected.  Positions and transverse
profiles of {\sc em} cluster showers are determined using the {\sc ces}
detector. Eleven strips (wires) around the most energetic strips
(wires) are grouped to form a {\sc ces} cluster.
To avoid spurious clusters made up by noisy channels, at least two
strips (wires) in each cluster are required to be above threshold. This
solution is more efficient than just requiring one strip (wire) with
high energy, as was done in Run I~\cite{RunIPhotonPRD}.  The
precise position of an {\sc em} cluster is determined using the
centroid of the most energetic {\sc ces} cluster inside the {\sc em}
cluster towers.  The position resolution achieved using this method is
about 2 mm for a single particle shower.
The {\sc CES} cluster centroid is also required not to be close to the
edges of the {\sc ces} where the detector is not fully efficient.  In
particular it has to be within 21 cm from the center of the tower in
azimuthal direction ($X_{CES}$) and within 9~cm $<|z|<$ 217~cm along
the beam direction ($Z_{CES}$).  A calorimeter cluster isolation
energy $E^{iso}_{T}$ is defined at this level as the total transverse
energy inside a cone of radius R=0.4 in $\eta-\phi$ space, centered
at the {\sc CES} cluster position, but excluding the cluster energy.  A
cut of $E^{iso}_{T}<1$ GeV is applied.  The profile of the cluster is
compared with a single {\sc em} particle profile as measured in test
beam and $\chi^2$s quantifying the ``similarity'' are formed in
both the azimuthal and longitudinal directions
~\cite{RunIPhotonPRD}. The average of these two $\chi^2$s are required
to be less than 20.  No explicit requirements on jets are implemented
at L3.  A summary of the trigger cuts is reported in
Table. 1.

\begin{table}[hbt!]
\begin{minipage}{\linewidth}
\begin{center}
\caption[Photon Triggers]{Summary of the requirements
    implemented in the trigger at different levels.  At level 1 and
    level 3 only cuts on photon related quantities are implemented. At
    level 2 requirements on hadronic clusters are present as well.}
\begin{tabular}{lc}
\hline
\multicolumn{2}{c}{{\bf Level 1}}\\
Trigger Tower $E_T$  & $>8$ GeV\\
Trigger Tower $E_{HAD}/E_{EM}$ & $<0.125$\\
\hline
\multicolumn{2}{c}{{\bf Level 2 - Photon Cuts}}\\
L2 {\sc em} Cluster $E_T$  & $>12$ GeV \\
L2 {\sc em} Cluster $E_{HAD}/E_{EM}$ & $<0.125$ \\
L2 {\sc em} Cluster $|\eta|$ & $<1.2$ \\
L2 {\sc em} Cluster $E_T^{iso}$ &  $<1.0$ GeV \\
\hline 
\multicolumn{2}{c}{{\bf Level 2 - Jet Cuts}}\\
L2 $\sum{E_T}$ & $>20+\ptg$ GeV\\
L2 Jet & $>1$ \\
L2 Jet $|\eta|$ & $<1.78$ \\
\hline
\multicolumn{2}{c}{{\bf Level 3}}\\
L3 {\sc em} Cluster $E_T$  &  $>12$ GeV \\
L3 {\sc em} Cluster $E_{HAD}/E_{EM}$ & $<0.05$ \\
L3 {\sc em} Cluster $E_T^{Iso}$  & $<1.0$ GeV \\
L3 {\sc em} Cluster $\chi^2_{CES}$ &  $<20$\\
L3 {\sc em} Cluster $|X_{CES}|$  & $<21$ cm \\
L3 {\sc em} Cluster $|Z_{CES}|$  &  $9<z<217$ cm \\
\hline

\end{tabular}
\end{center}
\end{minipage}
\label{tab:trg_summary}
\end{table}
\section{EVENT SELECTION\label{sec-EventSelection}}
The events selected online are processed offline taking into account
the updated calorimeter calibration, the tracker alignment constants,
and the measured beam position in the data. 
The primary vertex location is determined by iteratively fitting the tracks to a
common point. In case more than one vertex is reconstructed due to multiple
$p\bar{p}$ interactions in the same bunch crossing, the primary vertex of
the event is considered that whose
associated tracks have the highest sum of transverse energy.
The transverse energies are then
determined with respect to this interaction.  In the
following the offline event selection is described.

\subsection{Photon Selection}
To eliminate the cosmic ray contamination from the sample, the total missing
transverse energy~\cite{var-def} is required to be less than 80\% of the
transverse energy of the photon candidate.  The primary event vertex position
along the beam direction is required to be within 60 cm from the center of the
detector.  Only events with an {\sc em} cluster with $E_T>12$ GeV are
selected.  The cluster position determined in the {\sc CES} detector is
restricted to $|X_{CES}|<$ 17 cm and 14~cm $<|Z_{CES}|<$ 217~cm.  These
fiducial cuts ensure the {\sc em} shower is contained inside the {\sc ces}
detector boundaries, allowing an accurate reconstruction of its transverse
profiles.  The isolation cut applied at the trigger level is refined offline
where the transverse energy in a cone $R=0.4$ around the {\sc em} cluster,
calculated using the event vertex, is required to be less than 1 GeV excluding
the photon transverse energy. The photon energy is corrected in average for the
contributions of multiple $p\bar{p}$ interactions in the same bunch crossing
(pile-up events) and for the photon {\sc em} shower leakage into neighboring
towers.  In addition, the isolation requirement is reinforced by rejecting
photon candidates with a reconstructed track pointing to it.  Photons
converted into $e^+e^-$ pairs in the tracking volume or in the beam pipe,
about 14\% of all photons emerging from the interaction point, are also
rejected by this cut. The {\sc ces} shower shape is compared to the one
generated by a single {\sc em} particle profile with the same technique used
at L3. A similar $\chi^2_{CES}<20$ cut is thus applied.  Photon candidates
with a second {\sc ces} cluster inside the associated {\sc EM} cluster and
with energy above 1 GeV are also rejected to suppress the multi-photon
background. The efficiencies of these cuts in selecting prompt photons are
described in Section~\ref{sec-SelectionEfficiency}.

\subsection{Photon Background Subtraction\label{sec-PhotonSubtraction}}
\begin{figure}[!htb]
\begin{center}
\includegraphics[angle=0, width=9cm,clip] {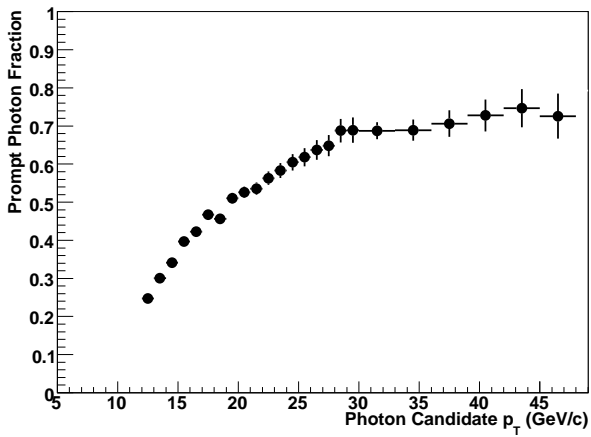}
\caption{Prompt photon fraction (number of prompt photon/number of
  candidate photons) in the data sample as a function of the photon
  candidate $p_T$. Only statistical errors are reported.}
\label{fig:pho_fraction}
\end{center}
\end{figure}

The photon candidates passing the above requirements are still contaminated by
multi-photons from neutral meson decay.  Two independent techniques are
employed to subtract this multi-photon background on a statistical basis. The
first one (``profile method'') exploits the difference in $\chi^2_{CES}$ of
the two components. Low $p_T$ prompt photons are expected to have a smaller
$\chi^2_{CES}$ than multi-photons which have a broader {\sc em} shower
profile. However, this method is not useful for {\sc em} clusters with
$p_T>35$ GeV: at such energies multi-photons are too collimated to produce
electromagnetic showers that are detectably broader than single photon.  The
second technique (``conversion method'') ~\cite{UA2-cpr} exploits instead the
different conversion probability of single and multiple photons when they pass
through the magnet coil, and it is approximately independent of $p_T$. Such
conversions are detected in the {\sc cpr} detector.  For both methods the
prompt photon content of the sample is given by:
\begin{eqnarray*}
N_{\gamma} =  \frac{\epsilon-\epsilon_b}{\epsilon_{\gamma}-\epsilon_b}\cdot N_{total},
\end{eqnarray*}
where $\epsilon_{\gamma}$ and $\epsilon_b$ are respectively the efficiencies
for prompt and multiple photons to pass a fixed $\chi^2_{CES}$ cut ({\sc cpr}
pulse height cut) in the case of the profile (conversion) method. Such
efficiencies are determined using both real data and simulated control samples
as detailed in ~\cite{RunIPhotonPRD}.  The number of photon candidates in the
sample is $N_{total}$ and $\epsilon$ is the fraction of these candidates
passing the cuts.  The two methods provide a consistent estimate of the prompt
photon content. In the following, for photon background subtraction, we
determine the prompt photon content using the profile (conversion) method to
photon candidates with $p_T<$ 35 GeV ($p_T>$ 35 GeV ). All the event
distributions, including $m_{jj}$, are accordingly weighted to subtract the
multiphoton background. The ratio of the number of prompt photons to the
number of photon candidates in the sample after the event selection is shown
in Fig.~\ref{fig:pho_fraction}.

\subsection{Jet Selection}\label{sec-jet_sel}
Hadronic jets are identified using an iterative cone clustering
algorithm \cite{JETCLU} with a cone radius R=0.7. Based on simulations of jet
fragmentation and of calorimeter response to hadrons the raw $E_T$ of the
jets are corrected for~\cite{CDF-JER-NIM}:
\begin{enumerate}[(i)]
\item the non-linear and non-uniform response of the calorimeter;
\item the undetected energy falling into uninstrumented regions of the
detector;
\item  the energy coming in average from multiple $p\bar{p}$ interactions
occurring in the same bunch crossing and the underlying event contribution;
\item the energy of low momentum charged particles that
  do not reach the calorimeter;
\item  the average energy loss due to particles falling  outside the
jet-clustering cone.
\end{enumerate} 
The jet corrections depend on the $p_T$ of the jet, its
pseudorapidity, and on the number of vertices in the event. They
amount, on average, to 25\%(15\%) of the jet energy for 15(50) GeV
jets.  In this analysis only events with two jets of $E_T>15$ GeV and
containing no additional jet with $E_T>10$ GeV are selected.  The
additional jet veto is introduced both to reduce the QCD background
and to improve the $W/Z$ dijet mass resolution by removing $W(Z)\gamma$
events with hard gluon radiation.

\section{SELECTION EFFICIENCY AND SIGNAL YIELD\label{sec-Systematics}}
In this section the trigger and offline requirement efficiencies in
selecting $\wzgqq$ signal events are calculated.  The trigger
efficiency is calculated for events satisfying all the offline
selection criteria. In turn, the offline selection efficiency is
evaluated using simulated $W(Z)\gamma$ events.

\subsection{Trigger Efficiency}
It is convenient to break up the trigger efficiency $\epsilon_{trg}$
in two components: 1) the photon selection efficiency
$\epsilon_{trg}^{\gamma}$ and 2) the efficiency related to hadronic
cluster requirements $\epsilon_{trg}^{jets}$ (see
Tab.~\ref{tab:trg_summary}).  

\begin{figure}[!htb]
\begin{center}
\includegraphics[angle=0, width=9cm,clip] {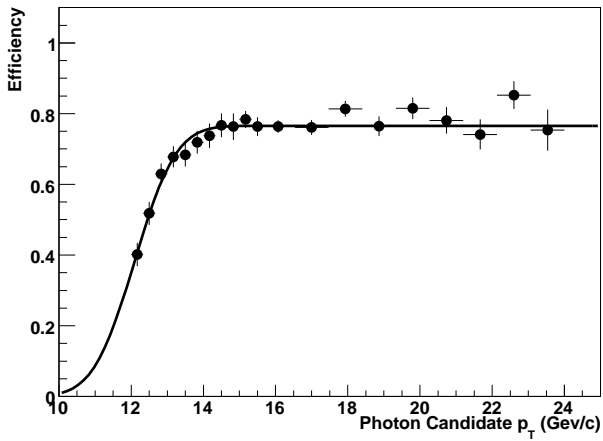}
\caption[\dps trigger efficiency]{Trigger efficiency for photon
  candidates as a function of $p_T$.
\label{fig:trg_eff}}
\end{center}
\end{figure}

The $\epsilon_{trg}^{\gamma}$ value is calculated as follows. First,
it is evaluated relative to a control sample collected by a trigger
with looser photon cuts (including a lower $p_T$ threshold). Then the
efficiency of this control sample is measured using a sample of
``unbiased'' photon candidates, {\em i.e.}  a sample where they have
not been used to trigger the data set.  The product of these two
contributions is shown in Fig.~\ref{fig:trg_eff}; this gives the
photon candidate trigger efficiency.  The value at the plateau
reflects the online/offline isolation energy differences while
the low $p_T$ turn-on is determined by the trigger threshold energy
smearing.
The final prompt photon trigger efficiency $\epsilon_{trg}^{\gamma}$ is
determined  by applying the photon background subtraction described in
Sec.~\ref{sec-PhotonSubtraction} to the  plot in
Fig.~\ref{fig:trg_eff}.  For prompt photons the plateau level increases to
85\% as they are more likely to pass the isolation cuts than the
multi-photons.

The $\epsilon_{trg}^{jets}$ is evaluated using the MC signal
sample described in Section~\ref{sec-SM-Prediction}.  The simulation
of the CDF calorimeter has been tuned to reproduce the response measured
in collider data~\cite{CDF-JER-NIM}.  The energies in the trigger
towers, the L2 $\sum E_T$, and the L2 jets quantities were estimated
using the online algorithms applied to the offline calorimeter tower
energies.  The accuracy of such estimates was checked in real data
against the actual online measurements, and for the quantities used in
the online selection the agreement was found to be within
1\%. The fraction of MC signal events passing the L2 jet
requirements
is  $\epsilon_{trg}^{jets}= 0.93$. 

In conclusion, the
combined trigger efficiency in selecting $\wzgqq$ event is
\begin{eqnarray}
\epsilon_{trg}= \epsilon_{trg}^{\gamma} \cdot \epsilon_{trg}^{jets} =0.76 \pm 0.01.
\end{eqnarray}

\subsection{Acceptance and Selection Efficiency\label{sec-SelectionEfficiency}}

The acceptance and efficiency of the offline event selection is estimated by
applying sequentially the cuts described in Section~\ref{sec-SM-Prediction} to
the MC signal sample.  In Table~\ref{tab:effsum} the offline cut relative
efficiencies, defined as the fraction of events passing a cut after having
passed all the previous cuts, are reported.  The MC simulation acceptance -
the fraction of generated events containing an {\sc em} cluster of $E_T>12$
GeV - reflects the choice of the $p_T$ photon generation cut (10 GeV). A lower
cut at the generation level is needed to avoid threshold bias brought about by
the finite detector resolution. The photon geometric acceptance includes the
pseudorapidity selection as well as the $X_{CES}$ and $Z_{CES}$ cuts.  The
accuracy of the efficiencies reported in Table~\ref{tab:effsum} depends upon
the precision of the detector simulation in reproducing the data.  The
electromagnetic particle response in the simulation is checked using electrons
from $Z\rightarrow ee$ and $W\rightarrow e\nu$ decays (a large sample of pure
prompt photons is not available in the data).  This comparison is used to
estimate the systematic uncertainties of the selection efficiencies.  
An account of these studies is given next:

\begin{enumerate}[(i)]
\item {$Z_{vertex}$ \em Cut}: The shape of the luminous region in real
  data was determined by fitting the vertex position in minimum bias
  events. The signal vertex position is simulated according to this
  distribution.  The fraction of events within $|z|<$ 60~cm in MC simulation
   matches the data within 0.5\%.
\item {\em Missing Energy Cut}: A change of the
  $E_T^{miss}/p_T^{\gamma}$ by 10\% resulted in a 2\% change in the
  selection efficiency, which is assigned as systematic uncertainty.
\item {{\sc had/em} \em Ratio}: The fraction of the unbiased electron
  from $Z$-boson decays which pass the {\sc had/em} cut in simulated
  and data events agrees within 1\%. We assume the same difference
  holds for photons, whose shower starts deeper in the calorimeter, and
  assigning a 1\% systematic uncertainty.
\item {\em Calorimeter Isolation:} The amount of energy surrounding a
  prompt photon {\sc em} deposition is determined by {\sc em}
  shower leakage outside the cluster and by underlying and multiple
  interaction events.  The accuracy of the simulation of the isolation
  cut measurement has been evaluated using cones of $R=0.4$ randomly
  placed in the photon fiducial region.  The energy collected in these
  cones can be considered an approximation of the isolation energy
  measured around {\sc em} clusters.  The fraction of such cones
  passing the isolation cut ($E_T<1$ GeV) in simulated $W\rightarrow
  e\nu$ events was found 3$\pm$2\% higher than in the data.  A correction
  factor 0.97 is applied to the MC isolation efficiency to
  account for the observed discrepancy. The 2\% uncertainty is
  included in the systematic errors.
\item {\em Track Isolation:} The track isolation efficiencies in
  simulated and real data events were found to be consistent within
  2\%.  The photon conversion $\gamma\rightarrow e^-e^-$ rate is used
  to tune the detector simulation for the amount of material present
  in front of the calorimeter.  The uncertainty of the track isolation
  efficiency includes any remaining deficiency in the material
  simulation.
\item {\em {\sc ces} $\chi^2$:} Photon and electron {\sc em} shower
  profile are simulated using the information collected during the single
  electron test beam.  As a consequence, for the efficiency of the {\sc ces}
  $\chi^2$ cut, a very good agreement (within 0.2\%) is observed
   between simulated and real data $Z\rightarrow ee$ events.
\item {{\sc ces} \em Cluster Isolation:} The {\sc ces} cluster
  activity around $Z$-boson decay electrons in MC simulation was found to
  match the data at the level of 3\%.
\item {\em Jet Cuts}: To assess the uncertainty on the jet cut
  efficiency the jet energy scale of all jets is shifted by one
  standard deviation (about 8(4)\% for jets of
  $E_T=$15(50)~GeV~\cite{CDF-JER-NIM}). This results in a 7\% relative
  change on the selection efficiency that we set as systematic
  uncertainty.  This is the dominant source of systematics.
\end{enumerate}

\begin{table}[htb!]
\begin{center}
 \caption[Event Selection Efficiency]
{Summary of the event selection cuts and their relative efficiency.}
\label{tab:effsum}
\begin{tabular}{lc}
\hline
{\em Analysis Cuts} & {\em Efficiency} (\%) \\
\hline
MC Simulation Acceptance &  62.1$\pm$0.1 \\
Photon Geometric Acceptance &  60.7$\pm$1.0\\
$|z_{vtx}|<60$ cm & 96.1$\pm$0.5 \\
Missing $E_T$ Cut & 90.6 $\pm$2.0 \\
\hline
\multicolumn{2}{l}{{\sf Total Acceptance:} $A_{kin}=0.33 \pm 0.01 $}\\
\hline
{\sc had/em} Ratio & 94.5$\pm$1.0\\
Calorimeter Isolation & 80.8$\pm$2.0 \\
Track Isolation & 80.2 $\pm$2.0\\
CES $\chi^2$  & 99.6$\pm$0.2 \\
{\sc ces} Cluster Isolation & 94.8$\pm$3.0 \\
\hline
\multicolumn{2}{l}{{\sf Total Photon Identification Efficiency:} $\eph=0.58\pm0.05$}\\
\hline
Jet 3 $p_T<10$ GeV Cut & 54.6 \\
Jet 2 $p_T>15$ GeV Cut & 82.8 \\
\hline
\multicolumn{2}{l}{{\sf Total Jet Selection  Efficiency:} $\ejets=0.45 \pm 0.03$}\\
\hline
\end{tabular}
\end{center}
\end{table}
Combining all the contributions in Table~\ref{tab:effsum}
($A_{kin}$, $\epsilon_{ph}$, and $\epsilon_{jet}$) with the trigger
efficiency $\epsilon_{trg}$, our estimate of the total signal
selection efficiency is
\begin{eqnarray}
\epsilon=\epsilon_{trg}\cdot A_{kin}\cdot\epsilon_{ph}\cdot
\epsilon_{jet}=0.065\pm0.006.
\end{eqnarray}
\subsection{Signal Dijet Mass Distribution and Signal Yield\label{sec-SimpleAnalysis}}
\begin{figure}[!tb]
\vspace{0.0cm}
\begin{center}
\includegraphics[angle=0, width=9cm,clip] {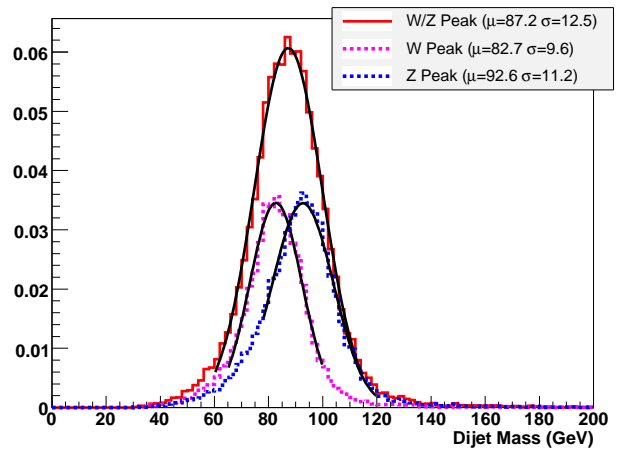}
\end{center}
\caption[Signal dijet invariant mass]{Normalized dijet invariant
    mass (solid line) distribution of the two leading jets of the
    selected $\wzgqq$ MC events. The individual contributions from the
    $W$ and $Z$ bosons (dashed lines) are shown. The fit results are
    reported in the inset. }
\label{fig:djmass_presel}
\end{figure}

The mass distribution of the two leading jet system for the simulated
signal events passing all the Table~\ref{tab:effsum} selection
criteria is reported in Fig.~\ref{fig:djmass_presel} along with the
individual $\gamma W$ and $\gamma Z$ contributions. Both the $W$ and
$Z$ mass distributions have non-Gaussian tails arising from initial
and final state gluon radiation. For the $Z$ we notice a larger low
mass tail due to the higher - on average - quark momenta compared to
the $W$ quarks.  In the range between 60 and 120 GeV the signal can be
adequately described by a single Gaussian with a mean value of 87.2
GeV and a width of 12.5 GeV. This shape is used to extract the
signal from the data.  The dijet mass resolution ($\Delta M/M$),
estimated by fitting a Gaussian function around the $W$ and $Z$ peaks,
 is 12\% for both gauge bosons. This is consistent with other MC dijet
mass resolution studies ~\cite{Higgs-Report}.  The expected number of
signal events in the sample is given by $N=\epsilon \times
\sigma_{\gamma W/Z}\times {\cal L}$, where $\epsilon=0.065$ is the
selection efficiency (without any mass window cuts), $\sigma_{\gamma
W/Z}=20.5$ pb is the SM cross section (reported in
Section~\ref{sec-SM-Prediction}), and ${\cal L}=184$ pb$^{-1}$ is the total
integrated luminosity of the sample. In the dijet mass window 60 $\leq m_{jj}
\leq$ 120~GeV, 227 signal events are expected among the 42462 events present
in the data. This corresponds to a signal over background ratio ($S/B$) of
$1/187$. For the current data set, the statistical significance - defined as
$S/\sqrt{S+B}$ - is 1.1.

In the following we show how the use of a
neural network in the selection process can substantially enhance the
sensitivity of the analysis.

\section{ADVANCED EVENT SELECTION\label{sec-NN-base}}

The basic event selection described in Sec.~\ref{sec-EventSelection}
has a rather mild discrimination power.  However, the 15 GeV jet
$p_T$ threshold cannot be increased since it would deplete the low end of the
dijet mass spectrum. Similarly,
 the rejection of the extra jet activity is meant more to improve
the dijet mass resolution rather than  suppressing the background.
Nevertheless, the kinematic and topological distributions of the final state
in signal and background events exhibit some differences that can
potentially be useful in enhancing the sensitivity of this analysis.
In fact:
\begin{enumerate}[(i)]
\item In signal events the $W(Z)$ boson has a low $p_T$ (since
  $p_T^{W(Z)}\sim p_T^{\gamma}$).  As a consequence the two jets are
  basically back-to-back with approximately the same energy, and the jet and photon
  directions are not correlated.  In contrast the dominant background
  ($\gamma+jj$ events) comes either from a $qg\rightarrow
  q\gamma$ production, where the quark balancing the photon radiates a
  gluon, or from a $qq$/$qg$ production, where one of the two outgoing
  quarks radiates a photon. In both cases, the radiated gluon/photon
  tends to be collinear with the radiating quark. Thus, the photon is
  either along or in opposite direction to the leading jet in the
  event.
\item In $\gamma+jj$ events the two leading jets are typically a
  quark and a gluon jet.  This is also true for  dijet production
  which is dominated at low $p_T$ by quark-gluon scattering.  For
  signal events instead the two leading jets are both quark jets.
\item The signal is characterized by the production of two colorless
  gauge bosons that constrains the initial and final state in a particular
  (color singlet) configuration.  The QCD background involves instead
  quarks and gluons with multiple color connections resulting in
  higher color radiation.
\end{enumerate}

\begin{figure*}[!htb]
\begin{center}
\includegraphics[angle=0, width=16cm,clip] {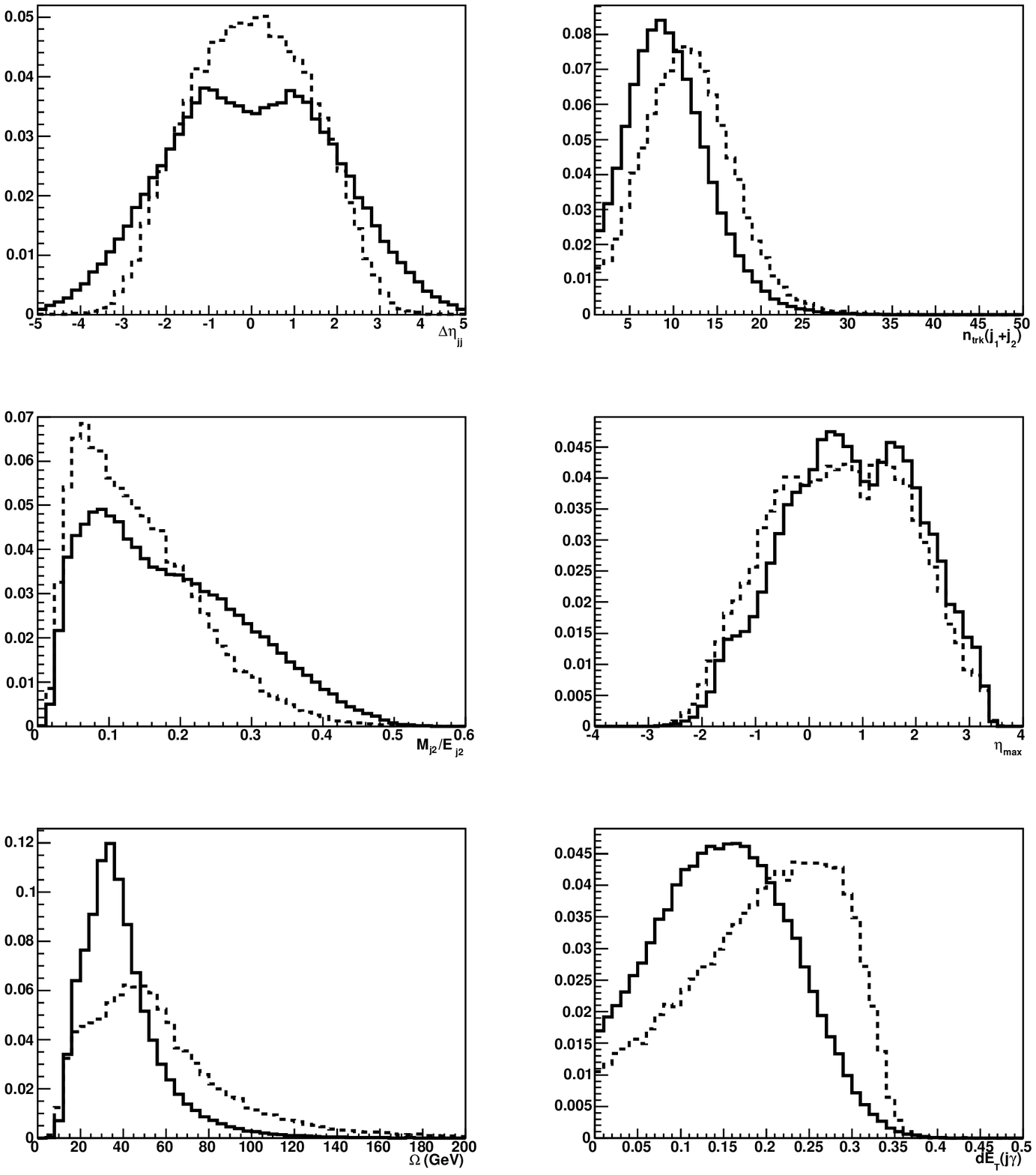}
\caption{Comparison of a few distributions for signal (dashed lines) and
  background (solid lines) events. All distributions are normalized to
  1. The observables reported are (from left to right, top to
  bottom): 1) $\Delta \eta_{jj}$, 2) $n^{j_1+j_2}_{trk}$, 3)
  $M_{j_2}/E_{j_2}$, 4) $\eta_{max}^{jets}$, 5) $\Omega$, 6)
  $dE_T^{j\gamma}$. A precise definition of these observables is
  provided in Table~\ref{tab:nn_table}.
\label{fig:nn-distribution}}
\end{center}
\end{figure*}

\noindent Hence, it is clear that the production of the signal and
the background events differs in many ways. However, it was not
possible to identify a set of selection criteria able to adequately
discriminate between signal and background while keeping an acceptable
signal yield. This is shown in Fig.~\ref{fig:nn-distribution} where the
signal and background distributions for a few observables are compared.
For these reasons we developed an artificial
neural network (ANN) selection to exploit subtle differences and variable
correlations. The ANN selection is applied to the events that have
already passed the simple kinematic cuts described in
Table~\ref{tab:effsum}.  The structure of the  ANN along with its
performance is described next.

\subsection{Neural Network Selection\label{sec-NN}}
In this analysis we employed the {\sc jetnet}~\cite{Jetnet} software
package to construct a feed-forward network~\cite{ForwardFeed}.  The
architecture of the network consists of one intermediate (hidden)
layer and a single output node.  For the network output $N_{OUT}$ a
target value of 1 for the signal and 0 for the background is
chosen. The training for the signal recognition is performed using as a
template $\gamma(W/Z)\rightarrow q\bar{q}\gamma$ events generated by {\sc
  pythia} (Sec.~\ref{sec-SM-Prediction}).  As background template
instead, a subsample of real data events is used. In fact it is not
trivial to simulate properly the QCD $\gamma+jj$ production because
of the interplay between the components associated to the hard process
(determined by matrix element calculations) and the components
generated by the development of the hard partons (described by parton
shower calculations)~\cite{Mrenna_matching}.  In addition, further
complications arise from NLO effects that cannot be neglected for an
accurate determination of the shape of the observable distributions
~\cite{Z_jets_NLO}, a key ingredient in an ANN training.  Considering
that less than 0.6\% of the data are signal events - based on
predicted production rates - data provide an excellent approximation
for background distributions. Only data events in the
60$\leq m_{jj} \leq$120 GeV signal mass window are considered in the ANN
training.



\begin{table*}[htb!]
\begin{center}
\begin{tabular}{cc}
\hline
{\em Property} & {\em Description}  \\
\hline
$\Delta \eta_{jj}$ & $\eta$ separation between the two leading jets  \\
\hline
$n^{j_1+j_2}_{trk}$ & Number of tracks inside a cone of size 0.5 \\ 
& in $\eta-\phi$ around the two leading jets \\
\hline
$M_{j_2}/E_{j_2}$ & Mass over energy ratio of the second jet ($M=\sqrt{E^2-P^2}$) \\
\hline
$\eta_{max}^{jets}$ & Maximum $\eta$ of the two leading jets \\
\hline
$\Omega$ & ``Intrajet Energy''  defined as $\Omega=\big(\sum E_T -
E_T^{jet1} - E_T^{jet2} -
E_T^{\gamma}\big)\big/{}\Delta L$  \\
 & where $\sum E_T$ is the $E_T$ scalar sum of the calorimeter towers  \\
 & in  the pseudorapidity region $(\eta^{DW}-0.3)<\eta<(\eta^{UP}+0.3)$   \\
 & and $\Delta L = \eta^{UP}-\eta^{DW}+0.6$ with  $\eta^{DW}=\min(\eta^{jet1},\eta^{jet2},\eta^{\gamma})$ \\
 & and $\eta^{UP}=\max(\eta^{jet1},\eta^{jet2},\eta^{\gamma})$. \\
 & The energies of the
photon and the two jets are {\em uncorrected} \\
\hline
$dE_T^{j\gamma}$ & $\big(E_T^{jet1}-E_T^{\gamma}\big)\big/\big(E_T^{jet1}+E_T^{jet2}+E_T^{\gamma}\big)$ \\
\hline
$\Delta \Phi_{jj}$ & Azimuthal angle between the two jets \\
\hline
$\max \Delta \Phi_{j\gamma}$ & Maximum azimuthal separation between photon and jets  \\
\hline
$\min \Delta \Phi_{j\gamma}$ & Minimum azimuthal separation between photon and jets \\
\hline
Sphericity & $S=3/2\cdot (Q_2+Q_3)$ with $0\leq S \leq 1$ \\
\hline
$\min \Delta \eta_{j\gamma}$ & Minimum $\eta$ separation between photon and jets \\
\hline
$\max \Delta \eta_{j\gamma}$ & Maximum $\eta$ separation between photon and jets \\
\hline
$\Delta \Phi_{\gamma W}$ & Azimuthal separation between the photon and the jet1-jet2 system \\
\hline
$\eta_{j_2}$ & Pseudo-rapidity of the second jet \\
\hline
$\Delta E_T^{jj}$ &  $E_T^{jet1} - E_T^{jet2}$ Transverse energy difference between jets \\
\hline
$\beta_W$ & $\beta$ of the jet1-jet2 system \\
\hline
Aplanarity & $A=3/2\cdot Q_3$ with $0\leq A \leq 0.5$ \\
\hline
$\cos\theta^{*}$ & cosine of the angle $\theta^{*}$ between 
the photon and the leading jet \\
&  directions calculated in the $\gamma$-jet reference frame \\
\hline
$\Delta \eta_{\gamma W}$ & $\eta$ separation between the photon and the jet-jet system \\
\hline
\end{tabular}
\end{center}
\caption
{Definition of the properties considered as input nodes for the neural
network. The sphericity and aplanarity are defined after~\cite{sphericity-def}.}
\label{tab:nn_table}
\end{table*}

\subsection{Variable Selection and Neural Network Training Tuning}\label{sec-NN_sel}
We consider a set of 19 input variables (or {\em nodes}) related to
the signal and background differences outlined above.  The selected
variables emphasize event and jet topologies, rather than absolute
kinematic values of the final state objects. This is done to preserve
as much as possible the shape of the $m_{jj}$ spectrum.  The list of
the ANN input nodes are given in Table~\ref{tab:nn_table} along with
their definitions.  In order to improve the performance of an ANN, it is
usually advisable to remove fully correlated variables from the set of
input nodes. To identify among our 19 variables the redundant ones, we
develop a ``ranking'' method that proceeds as follows.

 First, the most discriminating variable is determined by comparing the
 performance of 19 ANN's having each variable in
 Table~\ref{tab:nn_table} as a single input node.

\noindent The ratio $S/\sqrt{S+B}$ for a 
signal acceptance of 75\% is used as a figure of merit. 

Second, two input node ANN's are built. They have as a first input node the
variable found before and as a second node one of the remaining
variables.  The second best variable, defined as the property that
provides the best discrimination power when paired with the first
variable, is determined by comparing the significance of these ANN's.

\begin{figure}[!htb]
\vspace{0.0cm}
\begin{center}
\includegraphics[angle=0, width=9cm,clip] {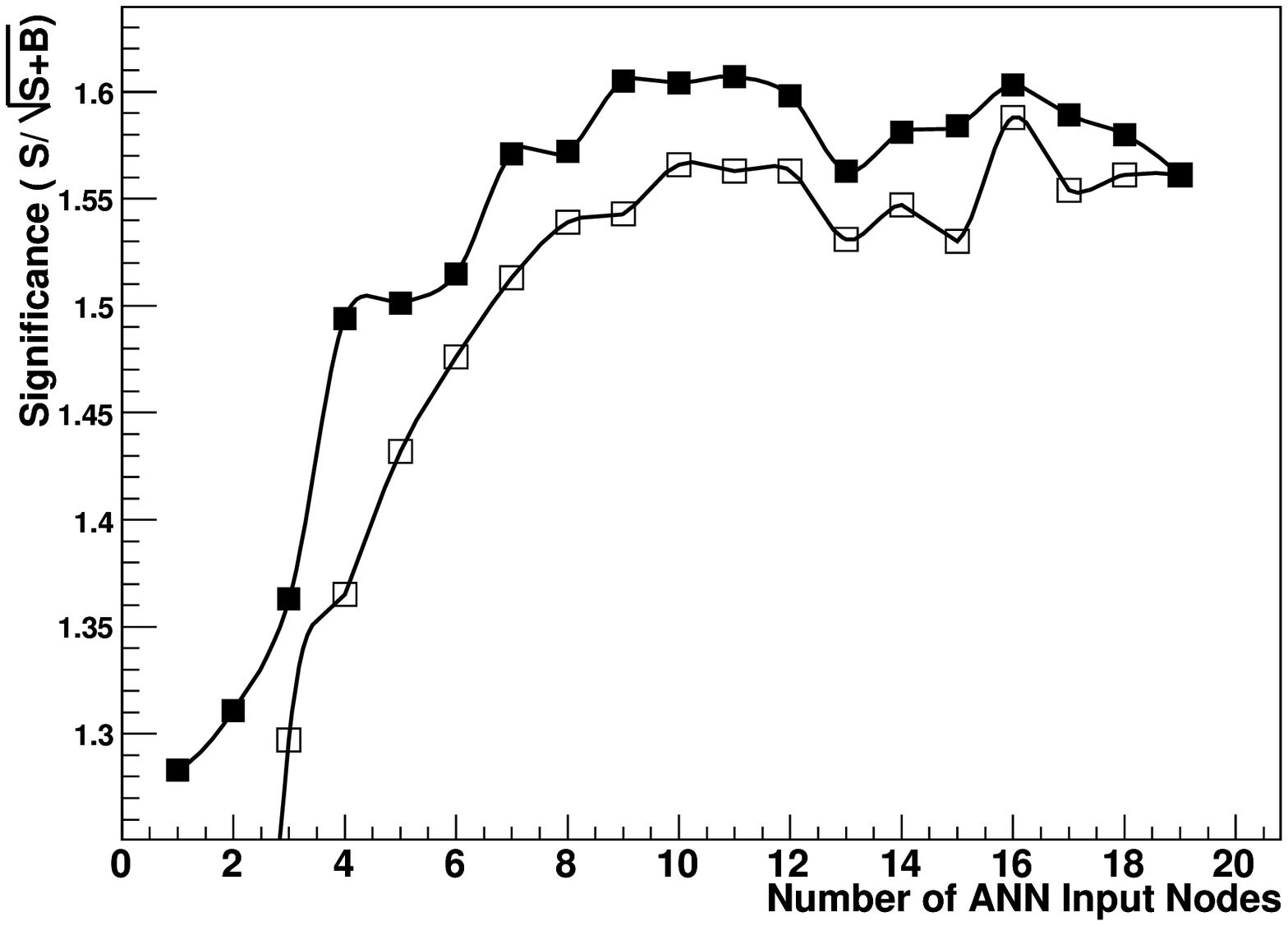}
\end{center}
\caption[$\nno$ distributions for the signal and for the background
  events]{Best (full square) and worst (empty square) ANN
    input variable combination in term of significance as a function
    of the number of input nodes for a signal efficiency of 75\%.
The fluctuations in the curves are due to small changes in the ANN
internal parameters one has to introduce when the number of inputs
increases. }
\label{fig:nn_ranking}
\end{figure}

The procedure is repeated, determining at each step the variable which,
in conjunction with the best set of variables found in the previous
step, forms the best performing ANN.  At the end, when all the
variables are considered, an ordered list of properties is generated.
In Table~\ref{tab:nn_table} the properties are listed in the order
resulting from this procedure.  The highest and lowest significance of
the ANN's built at the step $k$ ($k=1,..,19$) is shown in
Fig.~\ref{fig:nn_ranking}.  The ANN's discriminating power improves
with the number of input nodes until the properties that are
subsequently added become strongly correlated with those already
considered. At this point a plateau in performance is reached.  In our
case such a plateau appears at about $k=10$. Hence, only the first ten
properties listed in Table~\ref{tab:nn_table} are used as input nodes
in the final ANN.

The number of nodes, $N_{h}$, in the hidden layer is set to 17.
Several ANN's with $N_{h}$ from 11 to 30 were compared and no
significant differences in performance were observed.

\subsection{Neural Network Output and Improvement in Significance }\label{sec-NN_perform}
\begin{figure}[!htb]
\vspace{0.0cm}
\begin{center}
\includegraphics[angle=0, width=9cm,clip] {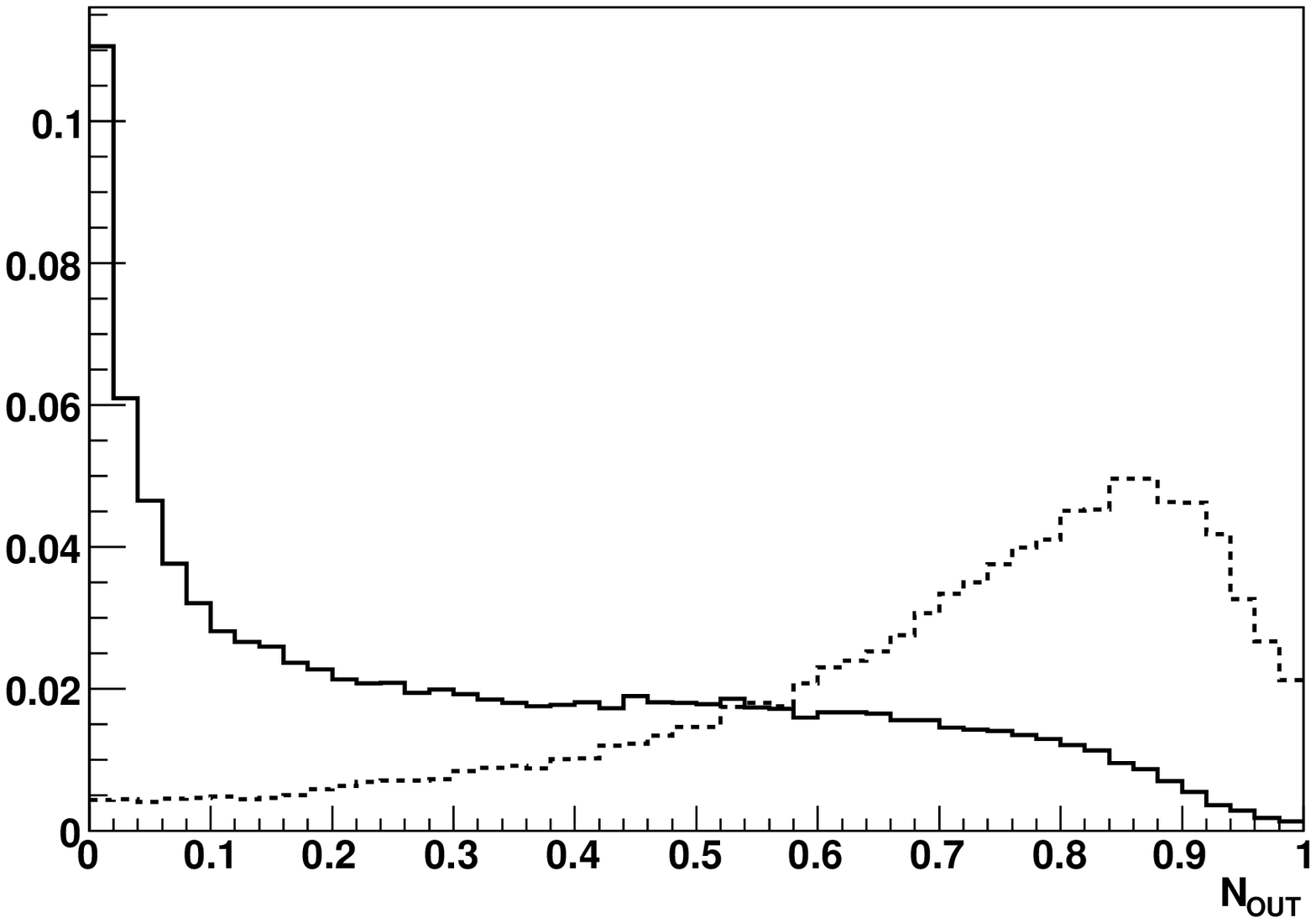}
\end{center}
\caption[$\nno$ distributions for the signal  and for the
  background (solid line) events]{The $\nno$ normalized distributions for the
  signal (dashed line) and for the background (solid line) events.}
\label{fig:nn_out}
\end{figure}

After the training, the ANN can be seen as a function associating a
real number $0.0 \leq N_{OUT}\leq 1.0$ to each event. The $N_{OUT}$
distributions for the signal and background samples are shown in
Fig.~\ref{fig:nn_out}.  Selecting events above some ANN output value
$N_{CUT}$ clearly enhances the signal sensitivity of the sample.

\begin{figure}[!htb]
\vspace{0.0cm}
\begin{center}
\includegraphics[angle=0, width=9cm,clip] {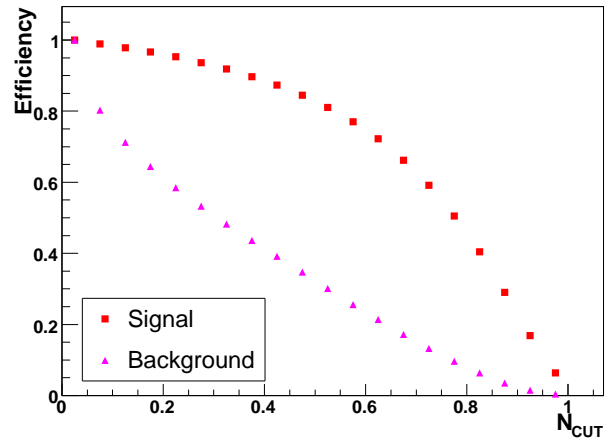}
\end{center}
\caption[S and B ANN Efficiency]{Efficiency for the signal
    and the background as a function of the ANN output threshold $N_{CUT}$.  }
\label{fig:nn_sbeff}
\end{figure}

In Fig.~\ref{fig:nn_sbeff} the signal (background) efficiency is shown
as a function of $N_{CUT}$.  In order not to deplete the signal yield
too much we set $N_{CUT}=0.6$. For this value the signal efficiency of
the ANN selection, $\epsilon_{NN}$, is 72\%.  After the ANN selection
the expected number of signal events is $S=164$, while 11691 data
events remain in the $60 \leq M_{jj} \leq 120$ GeV mass window. This
corresponds to an $S/B=1/71$ with a significance
$S/\sqrt{S+B}=1.51$, an improvement of 163\% (37\%) in
$S/B$ ($S/\sqrt{S+B}$) over the simple kinematic
selection reported in Sec.~\ref{sec-SimpleAnalysis}.  Moreover,
optimizing the size of the mass window, a significance of
$S/\sqrt{S+B}=1.86$ is obtained in the mass window $72 \leq
M_{jj} \leq 110$ GeV.
The data needed to achieve a significance of 5 is reduced by a factor
of two when the ANN selection built in this analysis is applied.

\subsection{Dijet mass spectrum}
After applying the $N_{OUT} > N_{CUT}=0.6$ cut to the data, the
starting point of the control region (at low $m_{jj}$) remains
approximately at the same value.
 This essential feature of our ANN
can be linked to the choice of having restricted the network training
sample to events with  $m_{jj}$ values within the signal region and
of not having explicitly used the energy of the two leading jets in
the ANN.  In addition, the ANN cut was applied to {\sc pythia} $\gamma+jj$ MC
events to check if some discontinuity was introduced in the $m_{jj}$
spectrum between the control and the signal region. As expected, the
$m_{jj}$ distribution was found to be very smooth over the entire
$m_{jj}$ range.

  As far as the $m_{jj}^{W/Z}$ signal distribution is concerned,
after the ANN selection, we observe no significant change in its
Gaussian shape with the same  mean and an improvement of about 1\%  in
resolution.  Hence, the ANN has similar selection
efficiency for $W$ and $Z$ boson events.

\subsection{Systematic Uncertainties on ANN Selection Efficiency}\label{sec-NN_sys}
Our final selection criteria are based on an ANN trained on simulated
and real data events.  Uncertainties in simulated quantities, such as jet
kinematic and topological properties, introduce an uncertainty in the
ANN selection efficiency.  The granularity of the CDF detector allows
an accurate determination of the directions of jets and photons. Thus,
the observables derived only from the directions ($\Delta \eta_{jj}$,
$\eta_{max}^{jets}$, $\Delta \Phi_{jj}$, $\max \Delta \Phi_{j\gamma}$,
and $\min \Delta \Phi_{j\gamma}$) rely only upon the final state
predictions made by the MC generator.  As discussed in
Sec.~\ref{sec-SM-Prediction}, a good agreement on final state
observables between the signal samples generated with {\sc pythia} and
{\sc madgraph} is found; thus systematic uncertainties associated with
these variables are negligible.

Other properties ($n^{j_1+j_2}_{trk}$, $M_{j_2}/E_{j_2}$, $\Omega$,
$dE_T^{j\gamma}$, and sphericity) rely on the accuracy of the CDF
detector simulation, in particular, on the calorimeter response to
particles and track reconstruction efficiency.  The calorimeter
simulation has been extensively tuned to real data using isolated
single tracks \cite{CDF-JER-NIM} while track reconstruction
efficiencies in data and MC are observed to be very similar.
The dominant uncertainty on these variables comes from the jet energy
scale.

A change of $1\sigma$ ~\cite{CDF-JER-NIM} in jet energy scale results
in a 27\% change in the combined $\epsilon_{jets}\cdot\epsilon_{NN}$
signal efficiency  value, which is assigned as total systematic
uncertainty on jet and ANN selection efficiency.

\subsubsection{Effect of Multiple  $p\bar{p}$ Interactions}
In the signal sample used to train the ANN
(Sec.~\ref{sec-SM-Prediction}) the contribution of additional $p\bar
p$ interactions (pile-up events) is not simulated.
In this data sample the average number of vertices is 1.7 and more than half
of the events contain at least one extra $p\bar p$ interaction. The ANN
variables were carefully chosen to avoid any bias from pile-up events.  The
jet energies already have soft interaction contributions subtracted
(Sec.~\ref{sec-jet_sel}), and only tracks coming from the primary vertex are
considered.  The only variable that could in principle be sensitive to
additional interactions is the intrajet energy $\Omega$ since it is made up
with uncorrected energies.  However, comparing data with single interaction MC
$\gamma+jj$ events, a difference of less than 4\% was observed for the mean
value of $\Omega$.\\ 
To gauge the size of a possible pile-up bias in our ANN, we divided the data
into two non overlapping sets: one containing events with only one
reconstructed vertex, and the second containing events with two or more
vertices. The ANN outputs for the two samples turned out to be very
similar. As a further  check a new ANN was built trained with these two
samples and based on the same ten variables used in our analysis.  With such a
training this new ANN is built to exploit any subtle (if any) pile-up
dependence of our input nodes and to discriminate events with one vertex from
events with more than one.  Similar $N_{OUT}$ distributions (within 1\%) were
observed in the two cases, showing that pile-up events do not have an
appreciable effect on our ANN.

\section{Background Estimation and $W/Z$ Peak Search \label{sec-Results}}
\begin{figure}[!htb]
\vspace{0.0cm}
\begin{center}
\includegraphics[angle=0, width=9cm,clip] {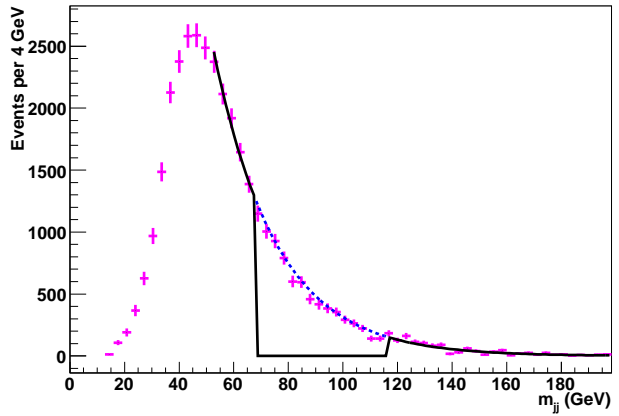}
\end{center}
\caption[Peak search final fit]{Dijet Mass distribution of
    the data after the NN selection cut. An exponential function
    $e^{P_0+P_1\cdot m_{jj}}$ is used to fit the two sidebands (solid line) and
    the result is interpolated inside the signal region (dashed
    line). The values $M_{min}$=52 GeV and
    $[M_{sig}^{L},M_{sig}^{H}]=[68,116]$ GeV are used to search for the 
    $W/Z$ mass peak.}
\label{fig:dj_final}
\end{figure}

\begin{figure}[!htb]
\vspace{0.0cm}
\begin{center}
\includegraphics[angle=0, width=9cm,clip] {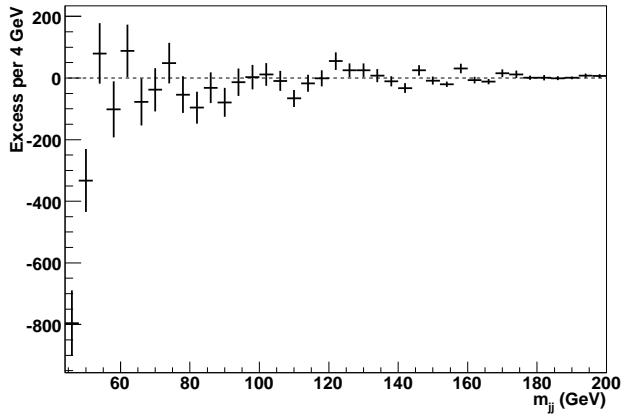}
\end{center}
\caption[Event Excess in the data]{Excess of events in the
    data with respect to the background prediction deduced from the sideband
    fits (bin errors do not include the background prediction
    uncertainties). The turn-on effect can be noticed in the first two bins
    (they are not included in the background fit). No evidence of any excess
    from the W/Z resonance production is found inside the signal region.  }
\label{fig:residual}
\end{figure}

The search for the W/Z peak is done by subtracting the background
contribution from the  data dijet mass distribution. The two control
regions are fitted with a smooth curve and interpolated inside the
signal region.  The functional form of the fit is provided by {\sc
  pythia} $\gamma+jj$ simulated events which are best described by
a simple exponential form $f(m_{jj})=e^{P_0+P_1\cdot m_{jj}}$. Hence,
the $m_{jj}$ spectrum from the data is fitted using  this function with
$P_0$ and $P_1$ as free parameters.  The fit is performed starting
from a minimum $m_{jj}$ value $M_{min}$ and excluding a mass window
$M_{sig}^{L} \le m_{jj} \le M_{sig}^{H}$ containing the signal region.
For reasonable variations of these three boundaries the changes in the
two fit parameters were found to be well within their statistical
uncertainties.  The fit parameters do not show any significant change
for values of $M_{min}$ greater than 52 GeV, while below that value we
observe a steep increase of the fit $\chi^2$ because of the  departure
of the dijet mass shape from an exponential behavior due to the
trigger threshold turn-on.  The fit using $M_{min}=52$ GeV,
$M_{sig}^{L}=68$ GeV, and $M_{sig}^{H}=116$ GeV is shown in
Fig.~\ref{fig:dj_final}. The interpolation within the signal region
(dashed line) is our estimate of  background. The dijet mass
spectrum after the background subtraction is shown in
Fig.\ref{fig:residual}.  A consistent result for the background estimate
was also found fitting only the high mass control region
($m_{jj}>M_{sig}^H$) and extrapolating  back inside the signal
region, but at the price of a 50\% larger uncertainty, confirming the
importance of the low mass control region for an accurate determination of the
background contribution. Since the subtracted distribution is
compatible with zero, we are not able to
identify a signal with the current data sample. In the next section we proceed to set an
upper limit on the $\gamma+(W/Z)$ production with the $W/Z$ boson
decays into hadrons.

\section{Cross Section Limit Calculation\label{sec-Limit}}

To extract the signal from the data a Bayesian-based statistical
procedure is applied. The region between 60 and 120 GeV of the
$m_{jj}$ distribution is divided into $N_{bin}=15$ bins, and 
the data events in each bin are regarded as a counting experiment
governed by Poisson statistics.  The total number of events expected
in the $i^{th}$ bin is $S_i+B_i$. The number of background events
$B_i$ is estimated from the dijet mass distribution as described in
Sec.~\ref{sec-Results}.  Since the stability of the control region fit
makes the error on $B_i$ very small, their values are held fixed. The
number of signal events are $S_i=\epsilon \sigma \lh s_i$, where
$\sigma$ is the cross section, $\epsilon$ the total selection
efficiency, $\lh$ the integrated luminosity, and $s_i$ the $i^{th}$ bin
content of the signal dijet mass density distribution as extracted
from the MC simulation (Fig.~\ref{fig:djmass_presel}).  At first 
 $s_i$ is held fixed  as well.  However, we show later how to take
into account the uncertainties affecting the shape of the signal
distribution.\\
The joint probability of measuring $n_i$ events when $\sigma \epsilon
\lh s_i+B_i$ are expected is given by

\[ P(n_i|\sigma,\ee,\lh) = \prod_{i=1}^{N_{bin}}
\frac{(\sigma \epsilon \lh s_i+B_i)^{n_i}}{n_i!}e^{-(\sigma \epsilon
  \lh s_i+B_i)}. \]

In Bayesian statistics the parameters $\sigma$, $\epsilon$ and $\lh$
are represented by probability distributions. Before the measurement
their corresponding {\em prior} density functions,
$\pi(\sigma),\pi(\epsilon)$, and $\pi(\lh)$, summarize our {\it a
priori} knowledge of them. Since no information on the cross section
is assumed before the measurement a uniform distribution is chosen as
its prior. In particular we define $\pi(\sigma)=0$ if $\sigma<0$ and
$\pi(\sigma)=1$ if $\sigma>0$.  For the efficiency and integrated
luminosity, we use the estimated values $\epsilon_0 \pm
\Delta\epsilon$ reported in Sec.~\ref{sec-SelectionEfficiency} and
$\lh_0 \pm \Delta \lh$ pb$^{-1}$ as reported in
Sec.~\ref{sec-overview}.  Their priors are assumed to be represented
by Gamma distributions $\gamma(x;\mu,\sigma_{\mu})$ with mean
$\mu=\epsilon_0$,$\lh_0$ and width $\sigma_{\mu}=\Delta \epsilon$,
$\Delta \lh$.
The expression for the joint {\em posterior} probability density for
$(\sigma,\epsilon,\lh)$ is provided by the Bayes' Theorem as:

\[ p(\sigma,\epsilon,\lh|n_i) = \frac{1}{\nh} P(n_i|\sigma,\epsilon,\lh)
\pi(\sigma)\pi(\epsilon)\pi(\lh), \]

\noindent where the normalization factor $\nh$ constrains the integral of
$p(\sigma,\epsilon,\lh|n_i)$ to unity when integrated over all the parameter
space.  To determine the cross section we calculate the marginalized {\em
posterior} probability distribution for $\sigma$ as:

\[ p(\sigma|n_i) =  \iint
p(\sigma,\epsilon',\lh'|n_i)d\epsilon'd\lh'.\] 

However, since the jet energy scale (JES) uncertainty results in a
change of the signal dijet mass distribution shape
(Sec.~\ref{sec-NN_sys}), $s_i$ cannot be considered fixed and its
dependence on JES systematics must be taken into account.  To include
this effect in the $p(\sigma|n_i)$ computation a new signal density
distribution is constructed moving the JES by one standard deviation.
Its bin content is defined as $s_i+\Delta s_i$.  As a consequence
the number of expected events is redefined as $\sigma \epsilon \lh
(s_i+t\Delta s_i)+B_i$, where the real number $t$ parametrizes the
uncertainty on the signal density shape. The prior density $\pi(t)$ is
assumed to be a Gaussian distribution centered at zero and with a width
equal to one.
The {\em posterior} density for $\sigma$,
including the new parameter $t$, is given by
\[
p(\sigma|n_i) = \frac{1}{\nh} \iiint P(n_i|\sigma,\epsilon',\lh',t')
\pi(\epsilon')\pi(\lh')\pi(t')\ d\epsilon' \ d\lh' \ dt'.
\]

As far as the cross section is concerned, this probability density
expresses the complete summary of the measurement. Upper limits (or a
central value with errors) can be hereby extracted from
$p(\sigma|n_i)$. The $p(\sigma|n_i)$ distribution was computed
numerically and no local maximum for $\sigma>0$ was found. 
The cross section upper limit $\sigma_{lim}$ at 95\% confidence level is
computed solving the equation:

\[ \int_0^{\sigma_{lim}} p(\sigma'|n_i)d\sigma' = 0.95. \]

\noindent It gives the value $\sigma_{lim}=54$ pb.

\section{Conclusions}\label{sec-conclusions}
We have developed a neural network approach to identify the dijet resonance of
the $W$ and $Z$ boson from the production of events having two jets with an
associated photon. As compared with a cut-based approach, the signal over
background ratio improves by 163\%, and the integrated luminosity needed for a
$W/Z\rightarrow jj$ peak to emerge from the huge QCD background is reduced by
a factor two. When applied to 184 pb$^{-1}$ of data collected by the CDF II detector,
 no evidence of a $W/Z\rightarrow jj$ peak is observed. The
standard model prediction for
$\sigma(p\bar{p}\rightarrow{}W\gamma)\times{}\mathfrak{B}(W\rightarrow{}q\bar{q'})
+
\sigma(p\bar{p}\rightarrow{}Z\gamma)\times{}\mathfrak{B}(Z\rightarrow{}q\bar{q})$
is estimated to be 20.5 pb for photons with $E_T>$ 10 GeV and
$|\eta|<1.2$.  A 95\% confidence level upper limit on this cross
section is extracted from the data with a full Bayesian approach and
found to be 54 pb.  The technique employed in this analysis can be
profitably extended to the search for small dijet resonance peaks
embedded in large multi-jet backgrounds.


\section{Acknowledgments}
We thank the Fermilab staff and the technical staffs of the participating
institutions for their vital contributions. This work was supported by the
U.S. Department of Energy and National Science Foundation; the Italian
Istituto Nazionale di Fisica Nucleare; the Ministry of Education, Culture,
Sports, Science and Technology of Japan; the Natural Sciences and Engineering
Research Council of Canada; the National Science Council of the Republic of
China; the Swiss National Science Foundation; the A.P. Sloan Foundation; the
Bundesministerium f\"ur Bildung und Forschung, Germany; the Korean Science and
Engineering Foundation and the Korean Research Foundation; the Science and
Technology Facilities Council and the Royal Society, UK; the Institut National
de Physique Nucleaire et Physique des Particules/CNRS; the Russian Foundation
for Basic Research; the Comisi\'on Interministerial de Ciencia y
Tecnolog\'{\i}a, Spain; the European Community's Human Potential Programme;
the Slovak R\&D Agency; and the Academy of Finland.



\begin{thebibliography}{24}


\bibitem{top-W-had}
  F. Abe {\em et al.} (CDF Collaboration), Phys. Rev. Lett. {\bf 80} 5720 (1998).
  A. Abulencia {\em et al.} (CDF Collaboration), Phys. Rev. D {\bf 73} 032003 (2006).
\bibitem{Higgs-Report} 
   L.~Babukhadia {\it et al.}  (CDF and D0 Working Group Members), FERMILAB-PUB-03-320-E.
\bibitem{UA2}
    J.~Alitti {\em et al.} (UA2 Collaboration),  Z. Phys. {\bf C49} 17 (1991).
\bibitem{CDF-lepton} 
    D. Acosta {\em et al.} (CDF Collaboration), Phys. Rev. Lett. {\bf 94} 041803 (2005).
\bibitem{Bocci-thesis} 
  A. Bocci, PhD Thesis, Rockefeller University, 2005.
\bibitem{Lum-error} 
    D. Acosta {\em et al.}, Nucl. Instrum. Methods A {\bf 494}, 57 (2002).  
\bibitem{CDF-RunIMarina}
    D. Acosta {\em et al.} (CDF Collaboration), Phys. Rev. D {\bf 73} 012001 (2006).
\bibitem{Pythia}
    T. Sjostrand \etal, Comp. Phys. Comm. {\bf 135} 238 (2001).
\bibitem{cteq5L}
    H.L. Lai  \etal (CTEQ Collaboration), Eur. Phys. J. {\bf C12}, 375 (2000).
\bibitem{var-def} CDF uses a cylindrical coordinate system in which
  $\theta$ and $\phi$ are the polar and azimuthal angles respectively,
  defined with respect to the beam direction $z$. Transverse quantities
  such as transverse momentum, $p_T$, are projections into the plane
  perpendicular to the beam direction. The pseudorapidity is defined
  as $\eta=-\ln{\tan{(\theta/2)}}$. The missing transverse energy is
  defined as the magnitude of  $\sum_i{E_T^i \vec{n_i}}$, where
  $\vec{n_i}$ is a unit vector that points from the interaction vertex
  to the $i^{th}$ calorimeter tower in the transverse
  plane. Calorimeter towers clustered in jets are corrected by a jet
  energy correction factor (Sec.~\ref{sec-jet_sel}).
\bibitem{madgraph}
    F.~Maltoni and T.~Stelzer, J. High Energy Phys. {\bf 0302}, 027 (2003).
\bibitem{nlo_b1}
    U. Baur \etal, Phys. Rev. D {\bf 48}, 5140 (1993).
\bibitem{nlo_b2}
    U. Baur \etal, Phys. Rev. D {\bf 57}, 2823 (1998).
\bibitem{CDF-detector}
    F. Abe et al., Nucl. Instrum. Methods Phys. Res. A {\bf 271},
    387 (1988).
    D. Amidei et al., Nucl. Instum. Methods Phys. Res. A {\bf
    350}, 73 (1994).
    F. Abe et al., Phys. Rev. D {\bf 52}, 4784 (1995).
    P. Azzi et al., Nucl. Instrum. Methods Phys. Res. A {\bf
    360}, 137 (1995).
    The CDF II Detector Technical Design Report,
    Fermilab-Pub-96/390-E

\bibitem{UA2-cpr}
    J. A. Appel {\em et al.} (UA2 Collaboration),  Phys. Lett. B {\bf 176} 239 (1986).
\bibitem{RunIPhotonPRD}.
   F. Abe {\em et al.} (CDF Collaboration), Phys. Rev. D {\bf 48} 2998 (1993).
\bibitem{JETCLU}
   F. Abe {\em et al.} (CDF Collaboration), Phys. Rev. D {\bf 45}, 1448 (1992).
\bibitem{CDF-JER-NIM}
    A. Bhatti {\em et al.}, Nucl. Instrum. Methods A {\bf 566}, 2 (2006).
\bibitem{Jetnet}
   C. Peterson, T. Rognvaldsson, and L. Lonnblad, Comput. Phys. Commun. {\bf 81}, 185 (1994).
\bibitem{ForwardFeed}
   J. Hertz, K. Anders, and R. G. Plamer, {\em Introduction to the Theory of Neural Computation} (Addison-Wesley, Boston, 1991).
\bibitem{Mrenna_matching} 
  S.~Mrenna and P.~Richardson, J. High Energy Phys. {\bf 0405}, 040 (2004).
\bibitem{Z_jets_NLO} 
   J. Campell and R.K. Ellis, Phys. Rev. D {\bf 65} 113007 (2002).
\bibitem{sphericity-def} The sphericity and aplanarity are defined
  using the momentum tensor $S^{\alpha\beta}=\frac{\sum_i
    p_i^{\alpha}p_i^{\beta}}{\sum_i |\mathbf{p}_i|^2}$, where the
  summation index $i$ is taken over the two leading jets and the
  photon, and ($\alpha,\beta$)=($x$, $y$, $z$). By standard
  diagonalization of $S^{\alpha\beta}$ the three eigenvalues $Q_1 \geq
  Q_2 \geq Q_3$, with $Q_1+Q_2+Q_3=1$, are found.







\end{thebibliography}
\end{document}